\documentclass[9pt,
twoside
]{pnas-new}
\setboolean{displaywatermark}{false}
\usepackage{xcolor}
\templatetype{pnasresearcharticle}

\title{\textcolor{black}{Explosive electrostatic instability of ferroelectric liquid droplets on ferroelectric solid surfaces}}

\author[a]{Raouf Barboza}
\author[a]{Stefano Marni} 
\author[a]{Fabrizio Ciciulla}
\author[a]{Farooq Ali Mir}

\author[b]{Giovanni Nava}
\author[b]{Federico Caimi}

\author[c]{Annamaria Zaltron}

\author[d]{Noel A. Clark}
 
\author[b]{Tommaso Bellini}

\author[a]{Liana Lucchetti}

\affil[a]{Dipartimento SIMAU, Università Politecnica delle Marche, via Brecce Bianche, 60131 Ancona, Italy}
\affil[b]{Medical Biotechnology and Translational Medicine Dept., University of Milano, 20054 Segrate, Italy}
\affil[c]{Dipartimento di Fisica e Astronomia G. Galilei, Università di Padova, via Marzolo 8, Padova, Italy}
\affil[d]{Department of Physics, Soft Materials Research Center, University of Colorado, Boulder, CO, 80305, USA}

\leadauthor{Lucchetti} 

\significancestatement{
	In this work we show that when sessile droplets of the newly discovered
	ferroelectric nematic fluid phase are deposited on a ferroelectric
	solid substrate, they can become suddenly unstable and disintegrate
	though the emission of fluid jets. The instability is due to the
	coupling between the polarizations of the liquid and solid materials,
	which induces the accumulation of polarization charges on the
	droplet-air interface and thus to the buildup of a repulsion pressure
	that eventually overcomes the surface tension. This new kind of
	polarization-induced Rayleigh instability crucially depends on the
	unique combination of polarization and fluidity of the ferroelectric
	nematic and might provide the basis for novel electrohydromechanical
	applications.}

\authorcontributions{\textcolor{black}{
TB, LL conceived the experiments.
RB, FC, GN built the setup.
AZ provided the substrates.
RB, SM, FAM preformed the experiments.
RB, SM, FC analyzed the raw data.
NAC, TB, LL interpreted the results and wrote the manuscript.}}
\authordeclaration{\textcolor{black}{The authors declare no competing interests.}}
\correspondingauthor{Liana Lucchetti.\\E-mail: l.lucchetti@univpm.it \\ Tommaso Bellini.\\ E-mail: tommaso.bellini@unimi.it}

\keywords{\textcolor{black}{ferroelectric liquid crystal $|$ lithium niobate $|$ electrostatic instability}} 

\begin{abstract}
We investigated the electrostatic behavior of ferroelectric liquid droplets exposed to the pyroelectric field of a lithium niobate ferroelectric crystal substrate.  The ferroelectric liquid is a nematic liquid crystal in which almost complete polar ordering of the molecular dipoles generates an internal macroscopic polarization locally collinear to the mean molecular long axis.  Upon entering the ferroelectric phase by reducing the temperature from the nematic phase, the liquid crystal droplets become electromechanically unstable and disintegrate by the explosive emission of fluid jets.  These jets are mostly interfacial, spreading out on the substrate surface, and exhibit fractal branching out into smaller streams to eventually disrupt, forming secondary droplets.   We understand this behavior as a manifestation of the Rayleigh instability of electrically charged fluid droplets, expected when the electrostatic repulsion exceeds the surface tension of the fluid. In this case the charges are due to the bulk polarization of the ferroelectric fluid which couples to the pyroelectric polarization of the underlying lithium niobate substrate through its fringing field and solid-fluid interface coupling. Since the ejection of fluid does not neutralize the droplet surfaces, they can undergo multiple explosive events as the temperature decreases.
\end{abstract}

\begin{document}

\maketitle
\thispagestyle{firststyle}
\ifthenelse{\boolean{shortarticle}}{\ifthenelse{\boolean{singlecolumn}}{\abscontentformatted}{\abscontent}}{}

\section*{Introduction}
\dropcap{T}he discovery of the ferroelectric nematic ({N\textsubscript{F}}) liquid crystal (LC) phase \cite{Chen:PNAS:2020} opens new possibilities in the study of the interactions between polar materials and electric fields. The fluid nature of this new phase combined with its polarity, makes its response to electric fields stronger and intrinsically different with respect to both ferroelectric solids and dielectric fluids. When in contact with solid substrates, the polarization $\boldsymbol{P}$ of {N\textsubscript{F}} is always parallel to the surface, no matter its chemistry, since any other direction would lead to an energetically costly accumulation of surface charge $\sigma=\boldsymbol{P}\cdot\boldsymbol{u}$, $\boldsymbol{u}$ being a unit vector normal to the surface. It is thus of particular interest to investigate the behavior at the interface between a {N\textsubscript{F}} phase and a ferroelectric solid, a situation that can be studied by depositing sessile {LC} droplets on a ferroelectric substrate. Indeed, the coupling of the two polarizations at the interface may give rise to new effects on the wettability and droplet’s contact angle.\\
As ferroelectric solid, we chose lithium niobate ({LN}) crystals that were used, with no surface treatment, as substrates for RM734 sessile droplets. Droplets contact angle was measured as a function of the substrate temperature on cooling from the isotropic ({I}) phase into the nematic ({N}) and {N\textsubscript{F}} phases.\\
As the temperature {T} is lowered, the contact angle mildly decreases in the {I} and {N} phases, signifying a slight increase of the wettability. New phenomena are instead observed when entering the {N\textsubscript{F}} phase. The contact angle abruptly decreases, followed – by further cooling – by a sort of droplet explosion, i.e. the abrupt occurrence of a shape instability, characterized by the ejection of jets of fluid which branch out into smaller streams and eventually disrupt into new small droplets. As {T} further decreases, these secondary droplets explode on their turn.\\
We understand this behavior as an analog of the instability predicted by Lord Rayleigh \cite{Rayleigh} for charged conductive liquid droplets above a critical charge-to-volume ratio. This instability arises from competing electrostatic and surface tension forces and leads to formation of charge-carrying fluid jets that reduce the droplet electric charge. Rayleigh instability has been widely investigated \cite{Achtzehn:EurPhysJD:2005, Smith:JPhysChemA:2002, Richardson:PRSA:1989, Duft:Nature:2003, Oh:SoftMatter:2017, Krappe:PhysScripta:2018, Grigorev:TechPhys:2005} and appears in numerous applications that include sprays used in native mass spectrometry, manufacturing, inkjet printing and 3D printing \cite{Jung:Nature:2021, Xie:ChemEngSci:2015} and biomedical applications \cite{Almeria:JCollIntSci:2010, Almeria:JCollIntSci:2014, Malik:JCollIntSci:2016}. However, the precise fission mechanism, which includes the birth and retraction of jets, has not been yet experimentally observed because of its fast kinetics \cite{Consta:Arxiv}. Rayleigh instability has also been theoretically investigated in infinitely long liquid crystalline jets \cite{Fel:JExpThPhys:2004, Yang:SoftMatter:2014} but no observations were reported.\\
The instability of ferroelectric sessile droplets on LN surfaces combines features typical of the Rayleigh instability, such as ejection and retraction of jets and formation of secondary droplets, with features due to its being driven by polarization, such as the repeated instability, the molecular order within the jets and the T dependence of the explosions which reflects the T dependence of P in the {N\textsubscript{F}} phase. Additional features are related to the LC ordering, such as the coupling between birefringence and polarization, which offers a new tool to characterize jets, and the relevant viscosity, ranging between 0.05 and 3 $\mathrm{Pa{\cdot}s}$ depending on T \cite{Chen:LiquidCrystals:2022} and thus larger than that of liquids typically used in Rayleigh instability studies \cite{Achtzehn:EurPhysJD:2005, Smith:JPhysChemA:2002, Duft:Nature:2003}, which enables an easier access to the kinetics of the effect.

\matmethods{
The ferroelectric liquid crystal 4-[(4-nitrophenoxy)carbonyl]phenyl2,4-dimethoxybenzoate (RM734) was synthesized as reported in \cite{Chen:PNAS:2020}. Its structure and phase diagram are shown in Fig. \ref{Fig:1}a and b. The {N\textsubscript{F}} phase appears through a second order phase transition when cooling from the N phase. The spontaneous polarization $\boldsymbol{P}$ of RM734 is either parallel or antiparallel to the molecular director $\boldsymbol{n}$ and exceeds $\mathrm{6{\mu}C/cm^2}$ at the lowest T in the {N\textsubscript{F}} phase (see Fig. \ref{Fig:1}c)\cite{Chen:PNAS:2020}. 
The lithium niobate (LN) substrates used in this work are 900 micron thick z-cut crystals. Experiments were performed on undoped, bulk iron-doped and diffused iron-doped substrates \cite{Zaltron:2016, Zaltron:2012}. Bulk doped substrates contain 0.1\% mol. of iron, while the diffused-doped contain 0.3\% mol. iron confined in a thin surface layer on the order of 30$\mathrm{{\mu}m}$\. The bulk spontaneous polarization PLN of LN crystals along the [0001] z-axis is of the order of 70$\mathrm{{\mu}C/cm^2}$ and does not depend significantly on T in the explored range since its Curie T is much higher ($\approx$1140°C). The huge bulk polarization of LN does not however translate in a huge surface charge density because of very efficient compensation mechanisms at the z-cut surfaces, lowering the surface charge density of thermalized LN to only about 10-2$\mathrm{{\mu}C/cm^2}$ \cite{Sanna:JPhysCondMatt:2017}. When temperature variations are induced, the surface charge of LN can significantly increase because of the pyroelectric effect \cite{Kostritskii:JApplPhys:2008, Kostritskii:JApplPhys:2010, Ferraro:APL:2008}, a transient phenomenon observable during and shortly after the variation \cite{Kostritskii:JApplPhys:2008} and due to the slow free charge relaxation in LN. The pyroelectric coefficient of undoped LN is of the order of $\mathrm{10^{-4}C/m^2K}$ at room temperature \cite{Byer:Ferroelectrics:1972, Bonfadini:SciRep:2019}, and increases by one order of magnitude around 100°C for both undoped and iron-doped crystals \cite{Tesfaye:FerroLettSec:2004}. Given the temperature used in our experiments, dictated by the RM734 phase diagram, we can thus expect an induced surface charge density of the order of 1$\mathrm{{\mu}C/cm^2}$, for T variations of a few degrees ramped in a short time compared to the LN charge relaxation.
RM734 droplets diameter, measured with a calibration slide, ranges from $\mathrm{1.25\times10^3 {\mu}m}$ down to $\mathrm{2.5\times10^2{\mu}m}$. Droplets were deposited on bare LN substrates previously slowly heated up to 200°C, corresponding to the RM734 I phase. Measurements of the contact angle $\theta$ as a function of temperature were performed with the set-up sketched in Fig. S1. For these measurements the substrate temperature was decreased down to 90°C after droplets deposition, cooling LN from below. Polarized optical microscope (POM) observations both in bright field and between crossed polarizer were also carried out and videos of the droplets behavior on cooling were recorded with a rate of 25 frames per second. For these experiments LN substrates were placed in a small oven suitable for the microscope stage, heated slowly to 200°C before droplet deposition and then cooled from below to 80°C. When not stated otherwise, the cooling rate in our experiments was of the order of 0.1°C/s. 
}

\showmatmethods{} 

\section*{Results}
\begin{figure*}
\centering
\includegraphics[width=\linewidth]{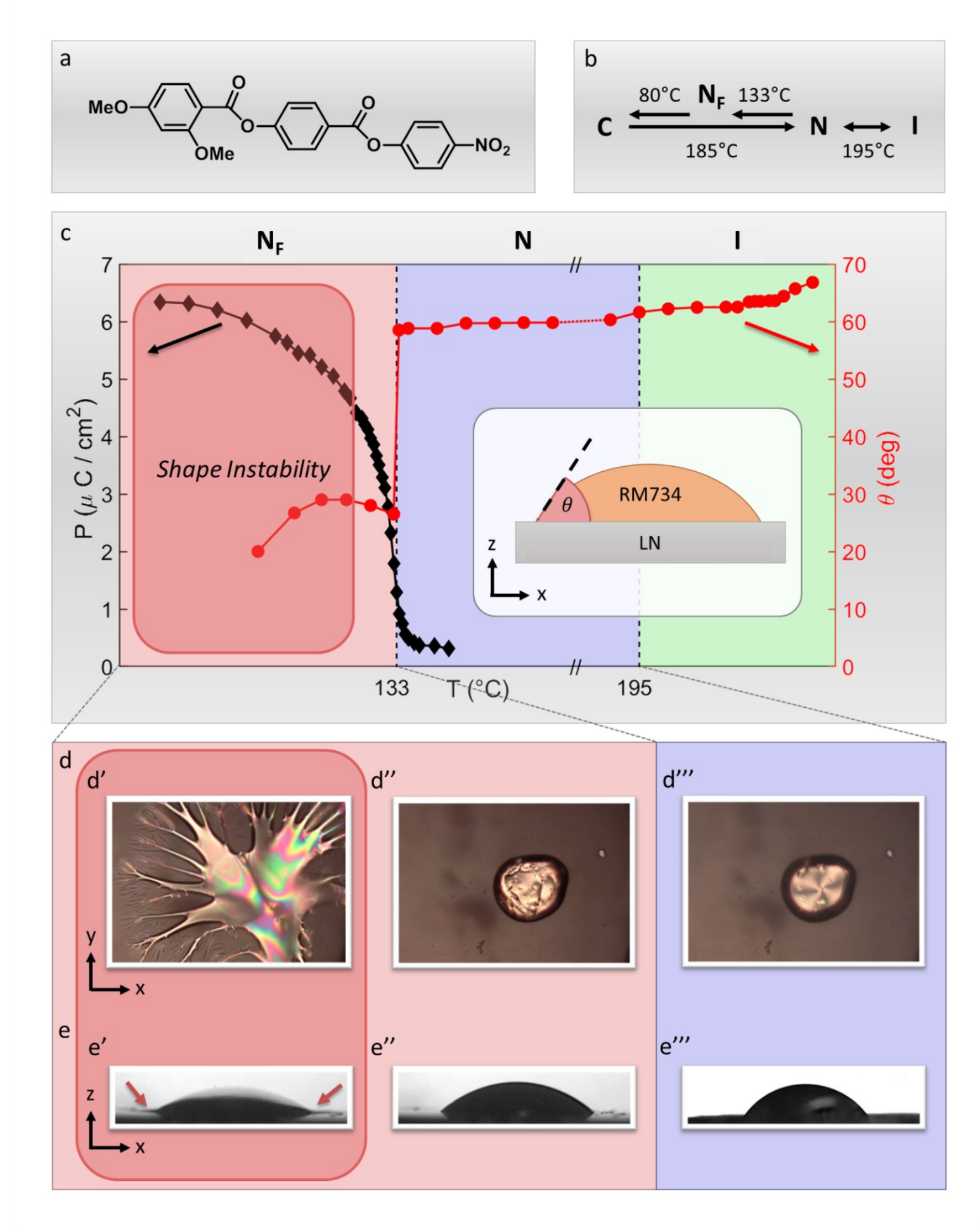}
\caption{a) and b) structure and phase diagram of RM734; c) spontaneous polarization $P$ (black curve) and typical behaviour of the contact angle $\theta$ on LN (red curve), as a function of temperature. The colored areas mark the different LC phases and the region of instability within the ferroelectric phase; d) POM images of a RM734 sessile droplet on an undoped substrate at three different temperatures: 160°C (d’’’), 130°C (d’’) and 109°C (d’), showing the droplet appearance between crossed polarizers in N phase (d’’’), the change in texture occurring upon entering the ferroelectric phase (d’’) and the explosive instability (d’). Droplet average diameter: $\mathrm{300{\mu}m}$; e) side view of the same droplet in the three situations. The decrease of the contact angle is evident in fig. 1e’’. The arrows in Fig. 1e’ indicate two fluid jets moving on the substrate.}
\label{Fig:1}
\end{figure*}

\begin{figure*}
\centering
\includegraphics[width=\linewidth]{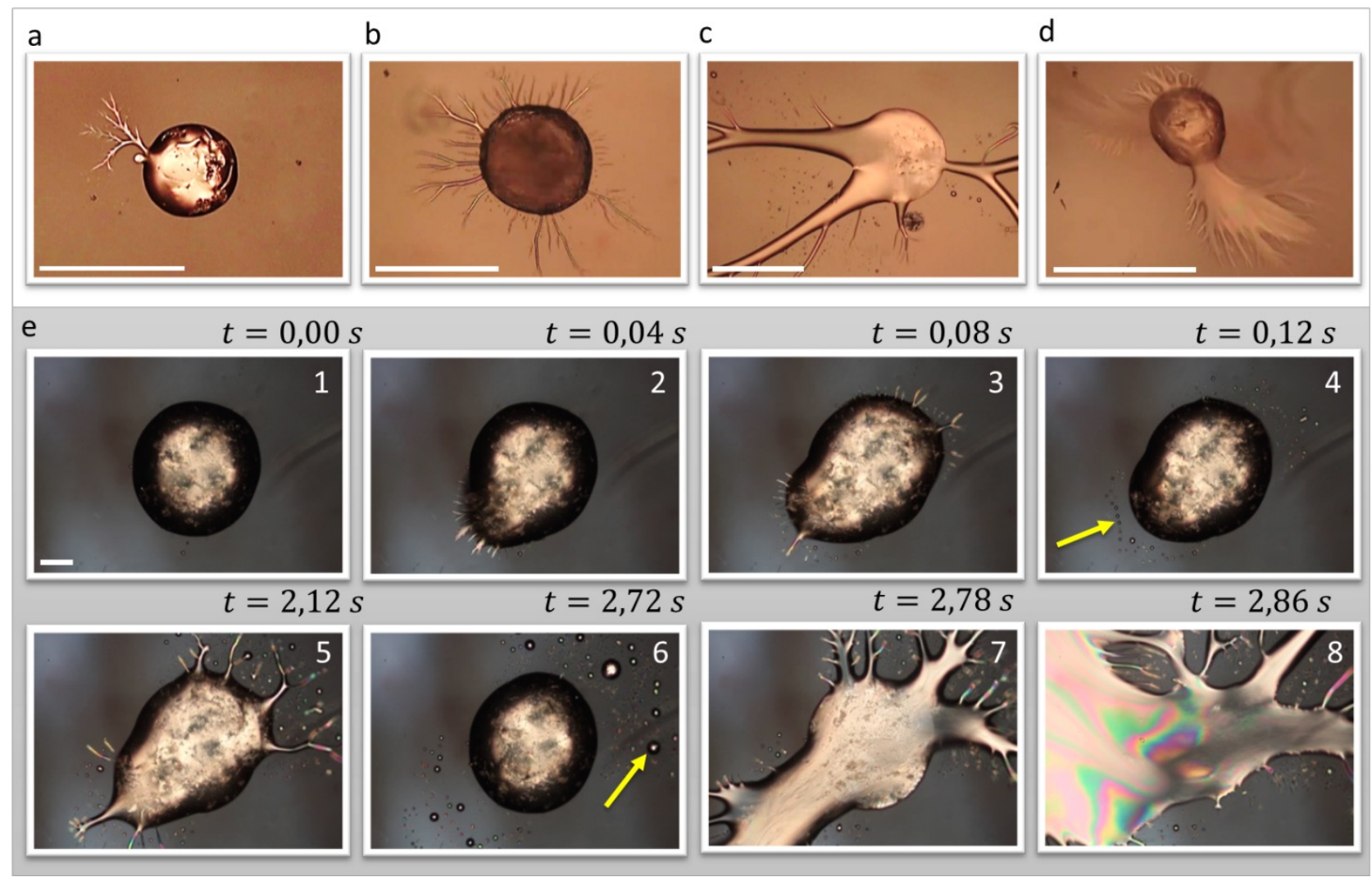}
\caption{Gallery of explosions, showing the initial stage of the phenomenon developing within the first tens of ms, for different droplets on iron-doped LN substrates (a-d). Droplets average diameter, temperature of instability and cooling rates: a) 300$\mathrm{{\mu}m}$, 128°C, 0.3°C/s; b) 480$\mathrm{{\mu}m}$, 104°C, 0.1°C/s; c) 500$\mathrm{{\mu}m}$, 131°C, 0.2°C/s; d) 250$\mathrm{{\mu}m}$, 124°C, 0.07°C/s; e) Snapshots extracted from Video S6 showing a RM734 droplet on an undoped LN substrate, in the temperature range \textcolor{black}{133-123°C}. The main features of the instability are observable: jet formation and branching at t = 0.04 s and t= 0.08 s (e2 and e3), formation of secondary droplets (e3 and e4), reformation of the mother droplet (e6), second and third instabilities (e5, e7 and e8). Droplet average diameter $\mathrm{1.1\times10^3{\mu}m}$, temperature of first instability T = 131°C, cooling rate 0.1°C/s.}
\label{Fig:2}
\end{figure*}

\begin{figure*}
\centering
\includegraphics[width=\linewidth]{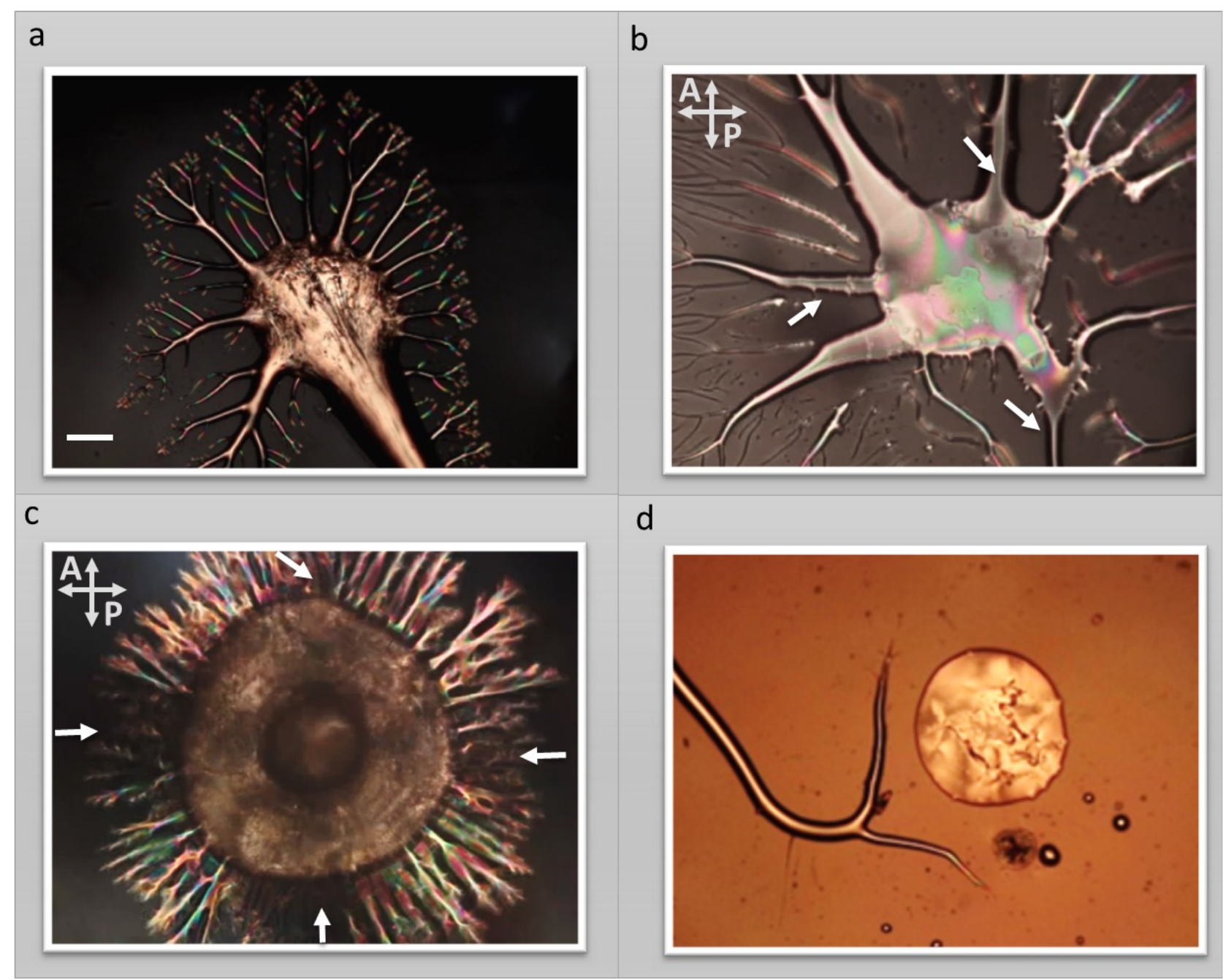}
\caption{a) LC jets ejected by  a droplet on an undoped LN substrate, branching to keep the largest possible distance; b) and c) RM734 droplets on undoped LN substrates, observed during jet ejection between crossed polarizer. Both pictures show that jets appear dark when parallel to the axis of one of the two microscope polarizer (white arrows), which demonstrates that both the LC director $\boldsymbol{n}$ and the polarization $\boldsymbol{P}$ are aligned along the jets main axis; d) LC jets on an iron-doped LN substrate deflecting in order to avoid RM734 droplets.}
\label{Fig:3}
\end{figure*}

\begin{figure*}
\centering
\includegraphics[width=\linewidth]{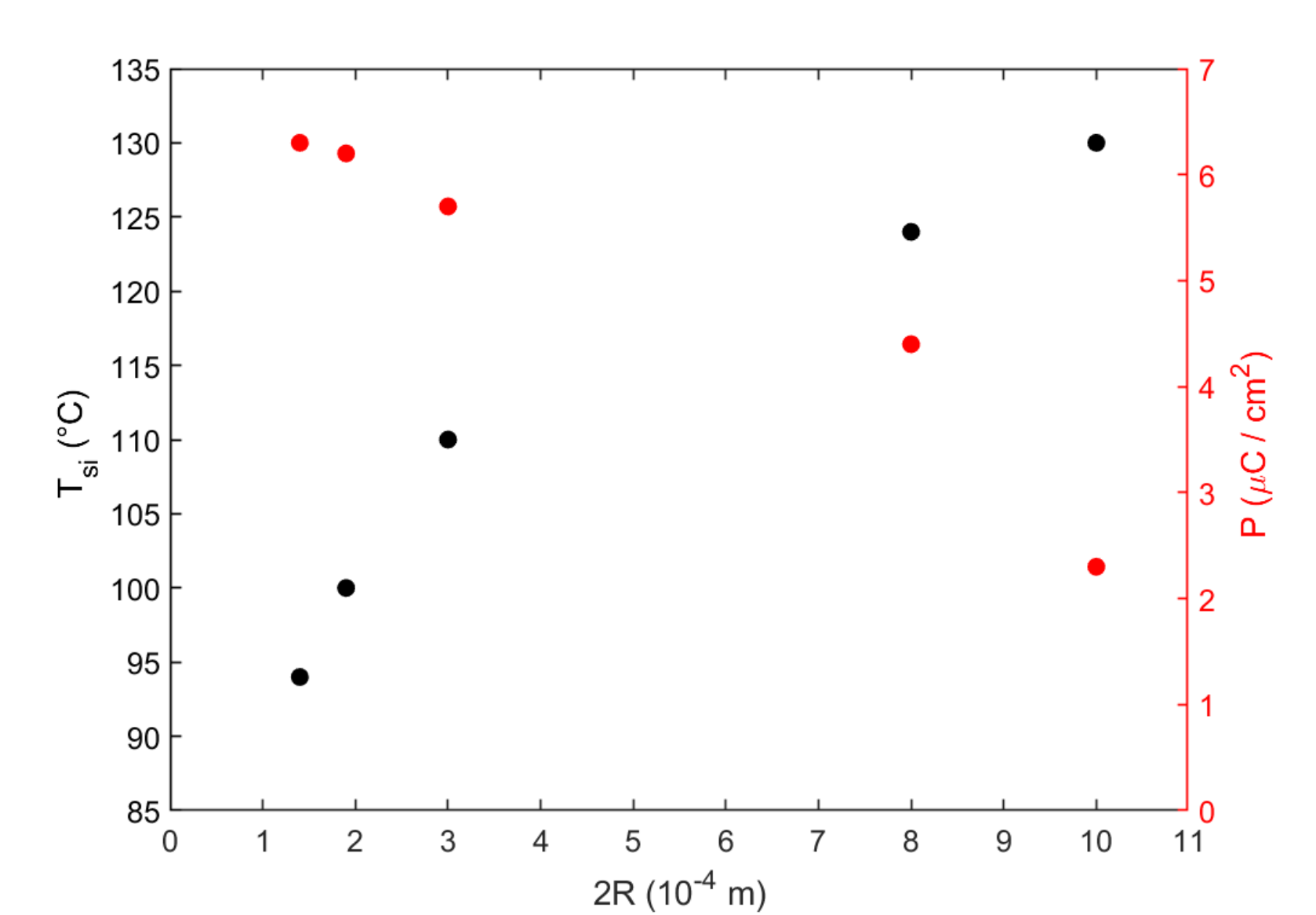}
\caption{Shape instability temperature $T_{si}$ (black curve) and spontaneous polarization $P$ (red curve), as a function of the average droplet diameter $2R$. The curve showing the dependence of $T_{si}$ on $2R$ has been built on the basis of a single experiment where several subsequent explosions were observed, in order to have consistent data not affected by different conditions such as the environment temperature, the specific LN substrate and the degree of cleanness of its surface. It shows that smaller droplets start ejecting fluid material at lower temperature thus requiring higher values of the spontaneous polarization, as suggested by the $P$ vs $2R$ curve. This indicates that small droplets are more stable than large ones.}
\label{Fig:4}
\end{figure*}

The typical behavior of the contact angle $\theta$ of RM734 sessile droplets on LN is reported in Fig. 1c as a function of T. A slight increase of the wettability upon cooling from I to N phase is observed, probably due to a decrease of the surface tension $\gamma$, generally observed in thermotropic nematics \cite{Gannon:PMA:1978}, and possibly also to the increased interaction with the charged LN surface, which tends to reduce the surface tension, and thus $\theta$, according to the Lipmann relation \cite{Mugele:JPhysCondMatt:2005}.\\
Upon entering the {N\textsubscript{F}} phase, we observe a sudden drop by half of the contact angle (Fig. \ref{Fig:1}c), as also apparent by visual comparison of the droplet profiles shown in Fig. \ref{Fig:1}e. Inspection of the droplet through crossed polarizers across the N-{N\textsubscript{F}} transition reveals that the large decrease of $\theta$ is accompanied by a change of texture. In the N phase the droplets show a well-defined director arrangement (Fig. 1d’’’), with $\boldsymbol{n}$ perpendicular to the solid substrate, as verified with thin cells built with two LN plates (data not shown) and parallel or slightly tilted to the air interface, forming a defect on the top, in analogy to what observed by Máthé et al \cite{Mathe:PRE:2022}. As the polar order develops, a growth in luminosity and the appearance of a more complex texture with topological defect lines is observed (Fig. \ref{Fig:1}d’’). The change in texture is compatible with the notion of a transition of the nematic director from perpendicular to parallel to the LN surface, in line with the general behavior of RM734 on solid substrates \cite{Caimi:SoftMatter:2021}.\\ 
As the droplets are further cooled into the {N\textsubscript{F}} phase, an explosive shape instability is observed, by which the droplet suddenly loses its typical dome shape and spreads on the LN substrate adopting transient complex geometries. This dramatic manifestation of the interaction between the ferroelectric LC and the ferroelectric substrate is observed at a temperature that, for fixed cooling rate, mainly depends on droplet’s size and, to a lower extent, to the specific LN substrate.  The appearance of these events upon entering the {N\textsubscript{F}} phase suggests that they are an electrostatically driven phenomenon, but they are driven solely by the {N\textsubscript{F}}/{LN} interaction, with no other sources of potential or electric field.  An example of such sudden explosion is shown in Fig. \ref{Fig:1}d’ and e’ (top and side views, respectively) characterized by a further flattening of the droplet and by the appearance of protrusions (indicated by arrows). We couldn’t observe significant texture changes within the droplets right before the explosion.\\
While droplet instabilities are invariably observed, the onset of the shape instability occurs with different morphologies, as reported in Fig. \ref{Fig:1}d’ and in Fig. \ref{Fig:2}a-d, both showing the initial stage of the phenomenon, developing within one video frame (40 ms), for different droplets. We observed violent smashes as the one reported in Fig. \ref{Fig:1}d’ (the video showing the whole phenomenon is reported in the SI as Video S1), slower ejection of single jets of fluid (Fig. \ref{Fig:2}a extracted from video S2), ejection of a great number of thin jets (Fig. \ref{Fig:2}b, from video S3), large jets (Fig. \ref{Fig:2}c, from video S4) and protrusion of large areas combined with jets ejection (Fig. \ref{Fig:2}d, from video S5). The different initial stages of the instability might be due to various factors including differences in LN substrates, in droplets average diameter, in cooling rates (see captions) and variations in the instability temperature. However, we couldn’t identify a clean connection between these factors and the observed morphologies.\\
After the initial burst, the instabilities develop with certain features common to all our observations. As an example, Fig. \ref{Fig:2}e reports a sequence of images extracted from Video S6 showing a RM734 droplet with $1.1\times10^3\mathrm{{\mu}m}$ average diameter on an undoped LN substrate, in the temperature range \textcolor{black}{133-123°C}. The initial ejection (in this case at T = 131°C) involves a deformation of the droplet shape (Fig. \ref{Fig:2}e1 and e2). The ejected jets appear to be themselves unstable, as indicated by the tendency of their tips to bifurcate (Fig. \ref{Fig:2}e3, e5, e7), giving rise to branched structures with several levels of ramification and forks characterized by a typical angle between branches of about 45°, an example of which is also shown in Fig. \ref{Fig:3}a. In the following stages of the instability, the explosion appears to lose its propulsion, the fluid velocity decreases and the jets undergo a fluid thread breakup type process, with the material in part moving back to the mother droplet (Fig. S2 and Video S3) and in part forming new droplets of smaller size (Fig. \ref{Fig:2}e4, e6 and S3). Overall, the instability causes the mother droplet to lose material leading to a temporary quiescent state. Upon lowering T further, new additional instabilities are observed, either involving the reduced-in-size mother droplet (Fig. \ref{Fig:2}e5 and e7), or the largest among secondary droplets. Indeed, as T decreases, several consecutive explosions of the original droplet (up to seven in a single experiment) can be observed enabling to determine the dependence of the shape instability temperature Tsi on the droplet size, as shown in Fig. \ref{Fig:4}. The RM734 polarization as a function of the droplet size is also shown in the figure. A comparison between the two curves indicates that smaller droplets start ejecting fluid material at lower temperature thus requiring higher values of \textcolor{black}{$\boldsymbol{P}$}, meaning that small droplets are more stable than large ones.\\
Observations under crossed polarizers of primary and secondary branches formed by the fluid jets, reveal that the director $\boldsymbol{n}$ is along their main axis, as indicated by the fact that jets parallel to either polarizer or analyzer appear dark (see Fig. \ref{Fig:3}b extracted from Video S1, and 3c). Since $\boldsymbol{P}$ is parallel to $\boldsymbol{n}$ in the {N\textsubscript{F}}, this observation implies that the jet tips are electrically charged, which suggests that electrostatic repulsion is at the origin of their tendency to bifurcate. This combines with various other features of the observed instabilities that clearly indicate the dominant role of electrostatics: the path adopted by the fluid jets; the spacing between the jets, and the fact that they only very rarely merge (see Fig. \ref{Fig:3}a); the extended distance at which inter-jet interaction can take place, which is too large for hydrodynamic or surface tension mechanisms; the interaction of jet streams with other droplets, which can be repulsive or attractive, as demonstrated by the sudden deflection of jets to either avoid (Fig. \ref{Fig:3}d) or collide with nearby droplets. \\
The role of electrostatics is also indicated by the fact that the instability here described crucially depends on the presence of the LN substrate. Sessile RM734 droplets are stable while cooling in the {N\textsubscript{F}} phase on other solid surfaces, including bare glass, Teflon and glass coated with various polymers. The instability appears identical on the two sides of the LN substrate, indicating that the sign of the charges of the LN surface that contacts the LC droplet is irrelevant. This finding suggests that solubilized ions are not relevant to the instability, which is likely entirely due to LC polarization. In our working conditions, jets always develop in contact with the LN substrate, as shown in Fig. \ref{Fig:1}e’. However, in modifying the set up such that the grounded metal oven lid is much closer to the LN surface, we could also observe tree-like jets developing vertically between the droplet and the lid (see SI, Fig. S5).\\
All the results here reported are obtained on cooling since the {N\textsubscript{F}} phase of RM734 is not obtained on heating. However, if a RM734 crystallized droplet is reheated, brief episodes of fluid jet ejection involving only a portion of the material are observed in proximity of the C-N phase transition, indicating that the {N\textsubscript{F}} ordering is transiently obtained even upon heating. \\
Temperature variations are essential to induce the instability. We find that instabilities are suppressed by holding the temperature fixed or by using much slower cooling rates even when the oven is left open, a situation in which strong but constant T gradients are present. These observations indicate that the charge density on equilibrated LN surface is not large enough to induce instability, and clearly suggest the essential role of LN pyroelectricity. We also noticed that when experiments are performed with the proper cooling rate but in conditions of increased T homogeneity, i.e. small T gradients, instabilities are still present but with slower jet ejection, an indication that charge inhomogeneities on the LN surface play some role in the jet kinetics, possibly through the appearance of in-plane field components.


\section*{Discussion}
\begin{figure*}
\centering
\includegraphics[width=\linewidth]{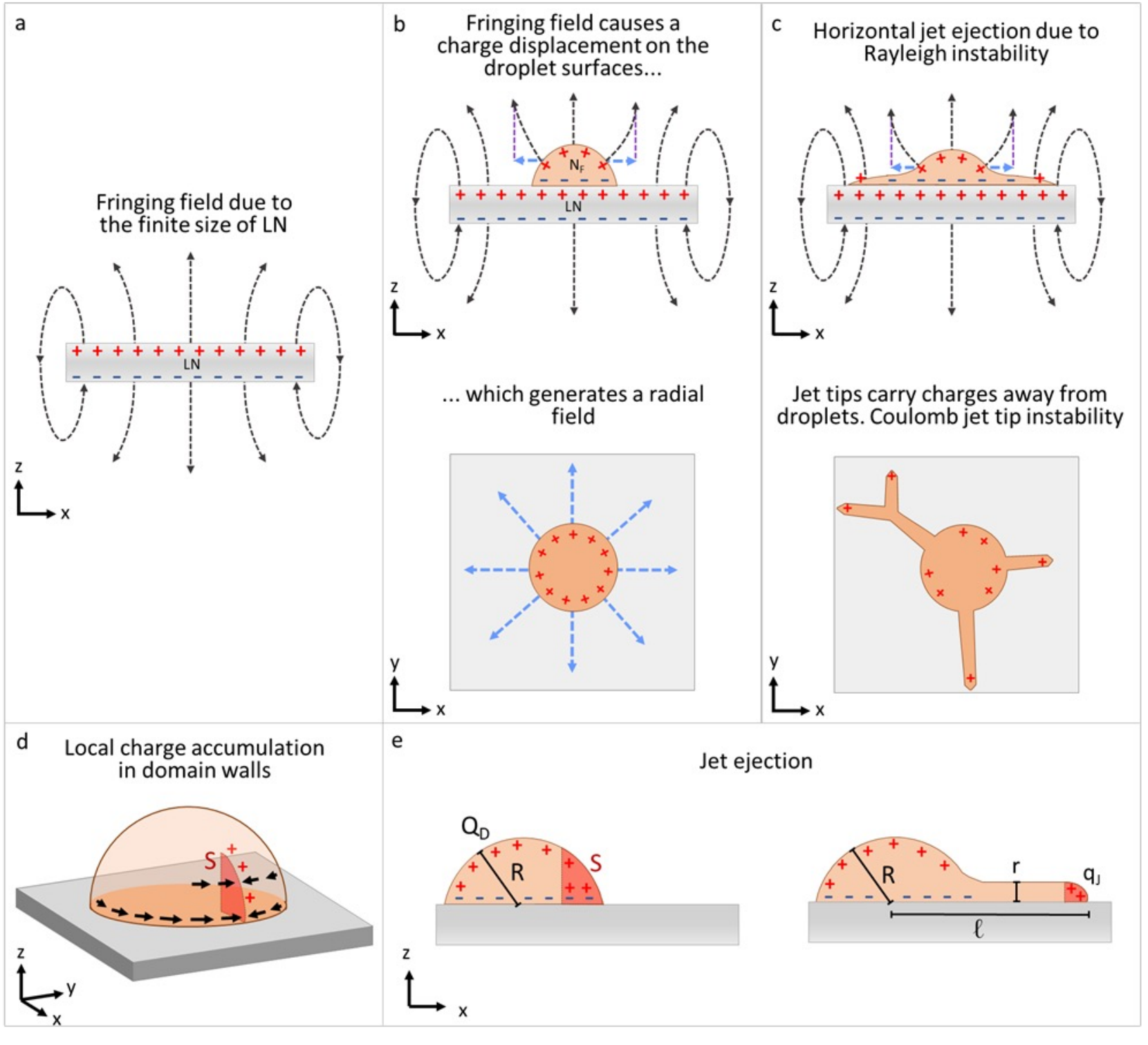}
\caption{Sketch showing the steps leading to droplet instability: the fringing field due to the finite size of the LN slab (a), combined to the interfacial coupling, produces an induced polarization of the ferroelectric droplet along z, which screens the field (b, side view). This polarization modifies the field surrounding the droplet that acquires a radial component (dashed blue arrows in b, side and top views) which drives jet elongation and motion at the instability (c). d) Illustration of a possible configuration leading to local accumulation of charges at the droplet rim and e) of the parameters used in the computation leading to both eq. (1) and the estimation of S}
\label{Fig:5}
\end{figure*}

\textcolor{black}{We interpret the explosive behavior of RM734 droplets on LN as an electrostatic instability, the closest example of which is represented by Rayleigh instability where droplets exhibit the sudden emission of fluid jets with increasing charging beyond a threshold}. Charged levitated droplets and electrosprays provide examples of this behavior, with the former showing that the jets form at a localized instability involving increased surface curvature and charge localization, in a system of positive feedback that results in the formation of surface tips and disruption into charged secondary droplets \cite{Duft:Nature:2003}.\\
In the present case this process occurs in the absence of free charges, but with the requisite charging within the droplet arising from the intrinsic polarization of the ferroelectric LC via its contact with the ferroelectric substrate. The LN crystal has finite size so its pyroelectric charging produces a fringing electric field $\boldsymbol{E}_{\mathrm{f}}$ external to the crystal, as sketched in Fig. \ref{Fig:5}a. The fringing field is a fraction $f$ of the internal field $\sigma_{\mathrm{LN}}/\epsilon_0$, with  $f\approx 10^{-3}$ depending on the crystal finite size \cite{Sloggett:JPhysA_MG:1986} and, in the region of interest, is largely in the vertical direction.\\
In the ferroelectric sessile droplet, the bulk polarization spontaneously self organizes to minimize the internal and external electric fields it produces. Generally, $\boldsymbol{P}$ will end up parallel to the droplet surface to avoid accumulation of surface charge and will adopt bend deformations which do not produce space charge, preventing nonzero $\nabla\cdot\boldsymbol{P}$ as much as compatible with geometrical constraints. Thus, we expect $\boldsymbol{P}$ to be nearly parallel to the LN-{N\textsubscript{F}} interface plane in the entire droplet. In the presence of the fringing field, which in this geometry is normal to the interface, the ferroelectric droplet becomes polarized. This takes place through a small reorientation of $\boldsymbol{P}$ by an angle such to deposit polarization charge on the droplet top and bottom surfaces that cancel the internal field. The displacement of polarization charges is of the order of $\sigma \approx f \sigma_{\mathrm{LN}}$  $\approx 10^{-3} \mathrm{{\mu}C/cm^2}$. The sign of the charges at the LN and air droplet interfaces is opposite and equal to the polarization charge of LN in the upper surface, respectively, as sketched in Fig. \ref{Fig:5}b.\\
Tests performed in geometries expected to yield larger Ef lead to more violent instabilities (see Fig. S4), confirming the relevance of the coupling mechanism between LN and ferroelectric droplet. {N\textsubscript{F}} polarization could be additionally promoted by the microscopic structure of the interface, leading to preferential alignment of the RM734 molecules with the dipoles of the solid substrate.
The amount of surface charge needed to cancel \textcolor{black}{$\boldsymbol{E}_{\mathrm{f}}$} within the droplet is much smaller, by orders of magnitude, than the spontaneous bulk polarization \textcolor{black}{$P$} of the RM734 {N\textsubscript{F}} phase (Fig. \ref{Fig:1}), indicating that very small reorientations of $\boldsymbol{P}$ are required to screen $\boldsymbol{E}_\mathrm{f}$ and that such screening takes place right at the transition to the {N\textsubscript{F}} phase. Since the Coulomb repulsion between the polarization charges deposited on the droplet surfaces is expected to effectively lower the surface tension, this notion agrees with the observation of the sudden decrease of the contact angle at the N-{N\textsubscript{F}} transition (Fig. \ref{Fig:1}) The jump in contact angle from about 60° to about 30° indicates a decrease of $\gamma$ of about 25\%, an estimate which can be obtained from the \textcolor{black}{Young-Dupré} equation \cite{Schrader:Langmuir:1995}. In this estimate we assume that the relevant variations in surface tension take place at the LC-air interface. A simple quantification of the charge density producing the observed decrease in surface tension can be given on the basis of the electrostatic energy $U$ of a uniformly charged disk, $U = \dfrac{8\pi}{3}k_C\sigma^2R^3$ \cite{Ciftja:ResPhys:2017}, where $k_C$ is the Coulomb constant and $R$ the disk radius which we take to be the same of the droplet. By assuming $\gamma\approx 3\times10^{-2}\mathrm{J/m^2}$ as for typical nematics \cite{Gannon:PMA:1978, Korjenevsky:LC:1993}, and thus $\Delta\gamma=10^{-2}\mathrm{J/m^2}$, we obtain $\sigma\approx2\times10^{-3}\mathrm{{\mu}C/cm^2}$, and a similar result ($\sigma\approx \mathrm{10^{-3}\mu C/cm^{2}}$) when a spherical or a half spherical shell are instead considered, a figure compatible with expectations. Being the fringing field essentially independent of T, we expect the resulting Coulomb effect on $\gamma$ to be T independent as well, as indeed observed (Fig. \ref{Fig:1}c). 
By becoming polarized, the droplet modifies the surrounding electric field, which acquires a planar (\textcolor{black}{$xy$}) radial component because of its dome shape (Fig.\ref{Fig:5}b). Such radial component appears to drive the motion of the jets at the instability. In our observations, jets start from the rim of the droplet, as in the sketch in Fig.~\ref{Fig:5}c, without involving, in the first stages of the ejection, its bulk. This appears to indicate that the edges of the droplets are likely location where topological defects more easily form and polarization charge accumulates, thus acting as trigger points for jet formation. In Fig.\ref{Fig:5}d we sketch a possible arrangement of the {N\textsubscript{F}} polarization that could lead to charge accumulation. To minimize charge accumulation, $\boldsymbol{P}$ must be collinear with the droplet rim. However, in the process of polar ordering, opposite directions of $\boldsymbol{P}$ might nucleate and converge in specific locations of the droplets, as in Fig. \ref{Fig:5}d. Indeed, the formation of complex domain walls and topological defects is always observed in {N\textsubscript{F}} droplets, as in Fig. \ref{Fig:1}d’’. The domain wall (red surface in Fig. \ref{Fig:5}d) is an area of charge accumulation $q=2PS$, $S$ being the wall area. The instability is produced when the Coulomb force $F_C=k_CqQ/R^2$, where $Q=\sigma A$ is the droplet charge and $A$ its surface, becomes larger that the force $F_S$ arising from the surface tension and opposing the formation of a cusp whose vertex is the localized charge pulling away from the droplet, which will be of order $F_S\sim\gamma\sqrt{S}$. In the initial instability reported in Fig. \ref{Fig:4}, $P \approx 2 {{\mu}C/cm^2}$. By assuming $\gamma \approx 10^{-2} N/m$ and $\sigma\approx 10^{-3}{{\mu}C/cm^2}$ , the condition $F_C>F_S$ leads in this case to $S > 0.2 {\mu}m^2$, an area which is much smaller than the droplet size, indicating that indeed the presence of even very small regions of local charge accumulation can be enough to initiate the jets.\\
As this condition is met, the instability turns to an explosive runaway process since the flow of the liquid ferroelectric in the nascent jet induces orientational order of the nematic director along the jet direction (Fig. \ref{Fig:3}b and c) which transports polarization charge to its tip in turn increasing the electrostatic repulsion. This can be recognized by assuming a simplified semicylindrical jet such as the one sketched in Fig. \ref{Fig:5}d and comparing the repulsive force $k_C\dfrac{Qq_t}{\ell^2}$ acting on jet tip with the one generated by the surface tension, $\gamma\pi r$.  $q_t=P{\pi}r^2/2$ is the charge of the jet tip and $r$ and $l$ the jet radius and length, respectively., we obtain that  $F_{Ct}>F_{St}$ when
\textcolor{black}{
\begin{equation}
P>\ 0.4\times10^{-5}\ \dfrac{\ell^2}{rR^2}\mathrm{{\mu}C/cm^2}
\end{equation}
}
a condition always verified in our experiments in which $\mathrm{0.13 mm}<R<\mathrm{0.6 mm}$ and $\mathrm{10{\mu}m}<r<\mathrm{100{\mu}m}$, $\mathrm{0.2mm}<\ell<\mathrm{3mm}$, as observed in Fig. \ref{Fig:2}. 

As they grow, jets follow paths that we understand as determined by a compromise among several factors as: interaction with the surface, local electric fields due to the presence of other droplets and jets, intrinsic instability given by the charge accumulation on their tip and possibly thermal gradients. The fact that the fluid jets run keeping a contact with the LN surface might result from a combination of the radial field generated by the polarized droplet and the minimization of energy cost for surface dilation, lower when in contact with the surface because of the smaller LN-RM734 vs air-RM734 surface tension. \textcolor{black}{Note that the increase of P upon decreasing temperature results in the local accumulation of larger values of polarization charges and to stronger repulsive forces}.\\
Careful observation of the instability (such as in the videos in the SI) reveals a wealth of additional intriguing phenomena, such as the polarity reversal in jets disconnected from the mother droplet, recursive jet pathways, attractive and repulsive secondary droplets, sudden collective instabilities. Part of these might reflect specificity of the LN surface and its thermal condition while others may be due to subtle combination of flow and polar ordering yet to be described. 

\section*{Conclusions}
The Rayleigh instability has been long known in conductive and dielectric liquid droplets when they host a sufficiently large free electric charge. Here we show that Coulomb shape instabilities can also be found in sessile droplets of the electrically neutral ferroelectric nematic liquid crystal when placed on a flat ferroelectric crystal. The coupling between the polarization in the solid and fluid materials induces the accumulation of surface charges on the droplet-air interface. As the polarization of the ferroelectric nematic grows by cooling the material, the local accumulation of polarization charges gives rise to repulsive forces that become unsustainable by the surface tension, \textcolor{black}{which instead prevails up to the onset of the instability}. The resulting instability takes the form of abrupt expulsion of polarized fluid jets whose electrically charged tips are repelled by the droplet. The tips themselves are unstable and often bifurcate, leading to a cascade of branched dynamic fluid jets and a whole new phenomenology of electrostatically dominated fluid motion.
The polarization-induced droplet electromechanic instability described here crucially depends on the properties of the newly discovered ferroelectric nematics. Our results show that the combination of polarization and fluidity not only opens the way to new electro-optic and electrokinetic phenomena, as widely acknowledged, but it also gives rise to body forces that, if controlled, might provide the basis for electrohydromechanical applications like soft robotics.

\acknow{
F. C., G. N. and T. B. acknowledge support by a PRIN2017 project from Ministero dell’Istruzione dell’Università e della Ricerca (ID 2017Z55KCW). N. A. C. acknowledge support by a 3,4, Grant No. and National Science Foundation Grants (DMR 1420736, DMR 1710711 and DMR \textcolor{black}{2005170}).
}

\showacknow{} 
\bibliography{references}


\end{document}



\maketitle


\section*{RM734 droplets preparation}
RM734 droplets were obtained following two steps. First of all a small amount of RM734 powder is deposited at room temperature on a clean glass slide and heated to 150°C to have a melt. To  create the initial droplets, a cold stainless needle is dipped into the melt and retracted, so that the droplet on its tip solidifies at contact with the surrounding air. To increase the size of the RM734 “pearl”, rapid (to avoid re-melting) successive dipping are performed. Then, the pearl is remolten into a droplet on the glass substrate. The glass slide is then cooled down to room temperature so that solidified droplets can be peeled off and reused on the proper substrate. The sizes of the droplets are controllable (size before peeling vs size of the remolten droplet) and are measured by means of a calibration slide. The average diameter ranges from 1.25 mm down to 250$\mathrm{{\mu}m}$.

\section*{Contact angle}
\begin{figure}[h]
\centering
\includegraphics[width=\textwidth]{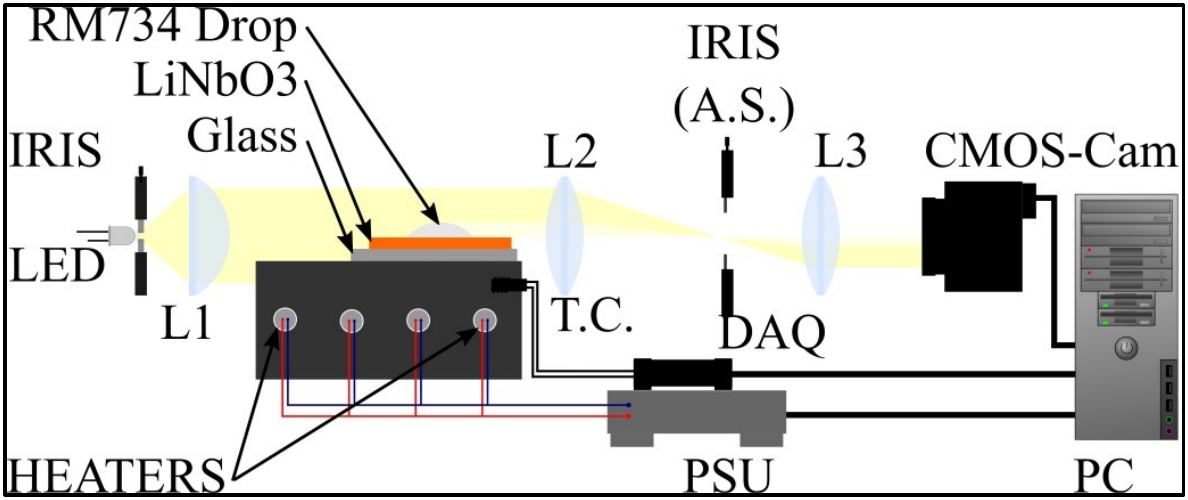}
\caption{Set up for measuring the contact angle of RM734 droplets on LN substrates. L1 collimating lens. L2 and L3 imaging and collecting lens. The first iris is used to reduce the width of the LED, the second as an aperture stop (A.S.). The CMOS sensor is placed in the focal plane of L3. The temperature of the LN substrate is controlled by a heating stage made of a slotted \textcolor{black}{aluminum} plate fitted with resistive heater cartridges powered by a programmable power supply unit (PSU). The setting point is precisely controlled (within 1.5°C) by feeding the measured temperature with a thermocouple (T.C.) and DAQ to a PID controller that drives the PSU.
}
\label{Fig:S1}
\end{figure}

Measurements of the contact angle as a function of temperature were performed with the set-up shown in Fig.~\ref{Fig:S1}, made of a collimated light source, an imaging system, and a hot stage. A white LED, an iris and a collimating lens are used to produce the collimated light source, while the imaging system is made of a collecting and an imaging lens with their focal back-to-back, an iris and a CMOS camera. Such a configuration provides the needed tele-centricity and a constant magnification ratio. To control the LC temperature, we use a slotted aluminium plate able to host resistive cartridges powered by a programmable power supply. The temperature of the hot stage is sampled with a thermocouple connected to a DAQ. The desired set point and the measured temperature are sent to a Labview PID controller that drives the programmable power supply.

\section*{Conventional Rayleigh Instability}
\textcolor{black}{The qualitative numerical evaluations proposed in the main text differ from the threshold of the classic Rayleigh instability of charged spherical droplets expressed through the so-called fissility parameter $X$. In that case, the increase of surface requires a break of spherical symmetry that does not apply to the sessile droplets, free to expand their interface with LN to compensate for the Coulombic pressure as in fact they do at the N-{N\textsubscript{F}} transition. Indeed, the} \textcolor{black}{threshold condition typically marking Rayleigh instability of a charged droplet of radius $a$ and surface tension $\gamma$ is achieved when its total charge $Q$ is such that the ratio $Q^2/64\pi^2\epsilon_0{\gamma}a^3$, defining the so-called “fissility parameter” $X$, is equal to 1. When expressed in terms of charge density at the droplet surface, $X$ takes the form:
\begin{equation}
X = \sigma^2 a/64\pi^2\epsilon_0\gamma
\label{Eq:SI}
\end{equation}
The condition $X = 1$ enables, via Eq. \ref{Eq:SI}, quantitative estimates of $\sigma_{\mathrm{RI}}$, the values of $\sigma$ necessary to yield the Rayleigh instability, as a function of the droplet size. We find that in the explored range, $\sigma_{RI}\mathrm{\approx 0.1{\mu}C/cm^2}$. This quantity should be responsible for the total screening of the fringing field $\boldsymbol{E}_f$ generated by the LN substrate, which, being $\boldsymbol{E}_f$ $\mathrm{\approx 10^3}$ times the internal field, should develop a surface charge density of the order of $\mathrm{10^3} \sigma_{\mathrm{RI}}\approx 100\mathrm{{\mu}C/cm^2}$, which is definitely too high to be pyroelectrically generated. On the other hand, the fissility parameter has been defined for a charged conductive spherical droplet, Rayleigh instability being in this case a discontinuous phenomenon where the loss of sphericity leads to an even less stable state and thus to a runaway effect by which initial lumps develop into fluid jets. Those described here are instead sessile droplets in contact with a solid substrate, they are not conductive but ferroelectric, thus they do not possess free charges but only polarization charges. The observed shape instability and jet ejection can be better understood as due to a local buildup of polarization charges, in turn due to the growth of the spontaneous polarization P around topological defects and constraints from the droplet boundaries, as discussed in the main text. When the electrostatic repulsion from the rest of the droplet due to this charge accumulation compensate the surface tension, a jet is started and the instability develops as described in the manuscript.}

\section*{Droplets images}

\begin{figure}[h]
\centering
\includegraphics[width=\textwidth]{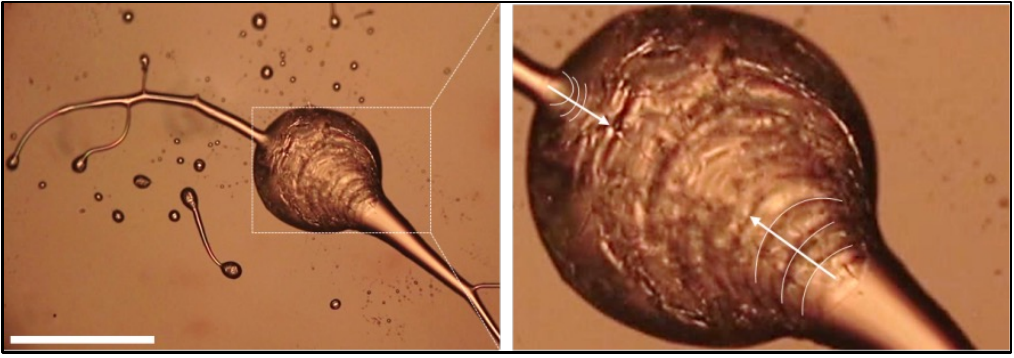}
\caption{RM734 droplet on a doped LN substrate. Two of the ejected jets are retracted from the droplet and carry back liquid crystalline material, as it is well visible in the blow up. Average droplet diameter: 450$\mathrm{{\mu}m}$. White bar = 500$\mathrm{{\mu}m}$. Extracted from Video S3.}
\label{Fig:S2}
\end{figure}

\begin{figure}[h]
\centering
\includegraphics[width=0.5\textwidth]{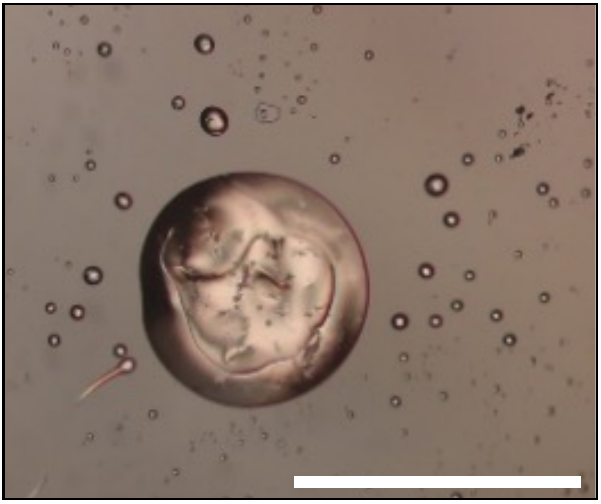}
\caption{RM734 droplet on a diffused doped LN substrate in a temporary quiescent state after jets ejection. Several secondary smaller droplets generated by the ejected jets are visible in the neighbourhood. Line defects in the mother droplet texture can also be observed. Average droplet diameter 400$\mathrm{{\mu}m}$. White bar = 500$\mathrm{{\mu}m}$.}
\label{Fig:S3}
\end{figure}

\begin{figure}[H]
\centering
\includegraphics[width=\textwidth]{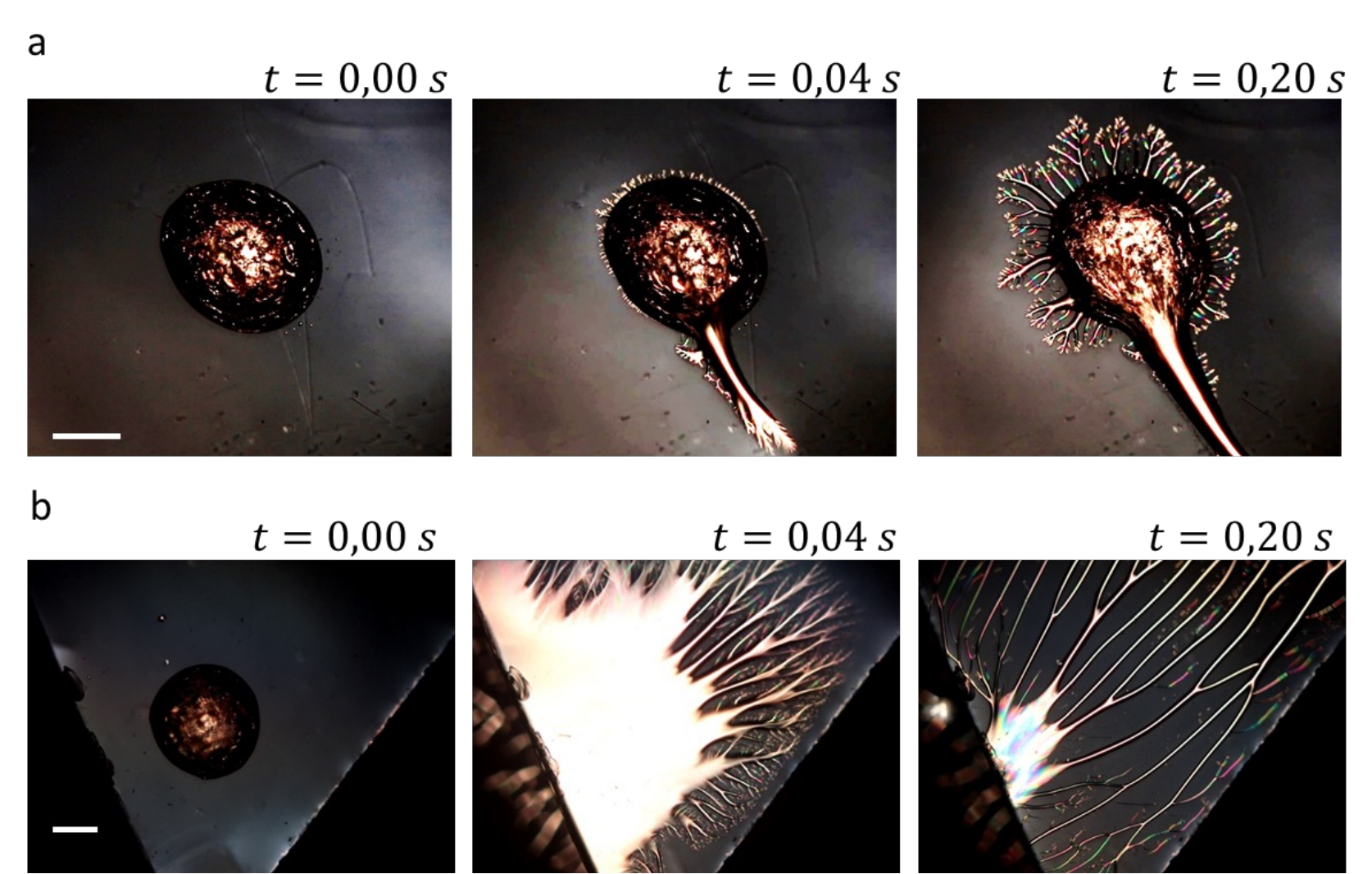}
\caption{RM734 droplets of similar size (average diameter = $1.2 \times 10^3\mathrm{{\mu}m}$) on an undoped LN substrate right in the centre (a) and close to the edges (b), at three different moments during cooling from T = 125°C. A comparison between the two figures shows that droplets instabilities develop with different strength, being more violent in case b) where the fringing field is larger. White bar = 500$\mathrm{{\mu}m}$.}
\label{Fig:S4}
\end{figure}

\begin{figure}[hb]
\centering
\includegraphics[width=\textwidth]{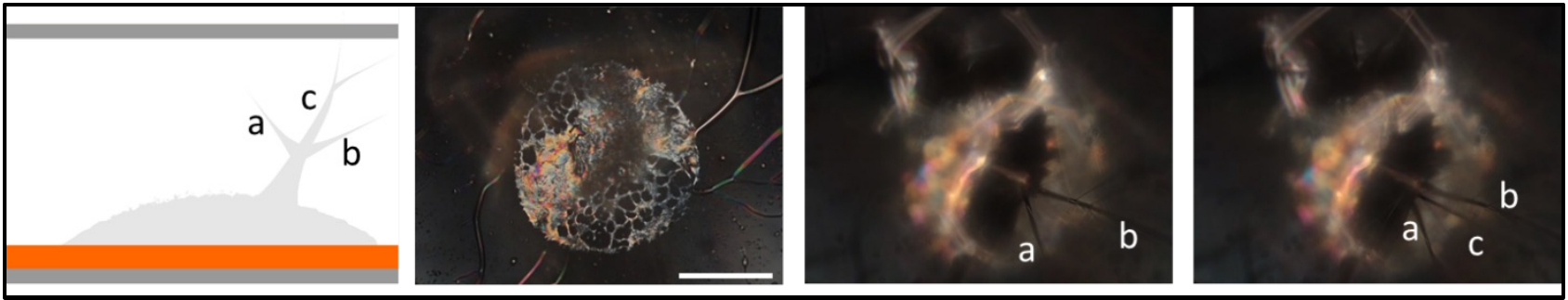}
\caption{Frames extracted by a video showing a rare event of 3D jet ejection (see main text). Three vertical jets are marked with letters. The vertical ejection occurs at 80°C, after several conventional 2D instability events. A sketch of the side view is also  reported. Undoped LN substrate. White bar = 500$\mathrm{{\mu}m}$.}
\label{Fig:S5}
\end{figure}

\newpage
\FloatBarrier
\movie{300$\mathrm{{\mu}m}$ RM734 droplet on a diffused doped LN substrate. In the first part of the video the substrate temperature ranges between 134 and 130° C and the transition from N to {N\textsubscript{F}} is observed at T = 133°C. The corresponding texture variation is clearly visible. In the second part, T goes from 125°C to 100°C and shape instability occurs at T= 109°C. Droplet shrinkage after explosion is also visible. In order to visualize the director rearrangement induced by the phase transition, the cooling rate was kept slow (0.01°C/s) in the first part and then faster (0.2°C/s) in the second part.}
\movie{320$\mathrm{{\mu}m}$ RM734 droplet on an undoped LN substrate observed under a polarised optical microscope. The substrate temperature ranges between 165 to 125°C. It is possible to observe the N/{N\textsubscript{F}} transition at T = 137°C resulting in a change of the LC texture with the appearance of defect lines that move as the temperature decreases. Droplet explosive instability starts at T = 128°C. Cooling rate: 0.3°C/s.}
\movie{450$\mathrm{{\mu}m}$ RM734 on a diffused doped LN substrate. Substrate temperature ranges between 125 and 60°C with a cooling rate of 0.1°C/s. Up to seven instability events are observed, the first occurring at T = 104°C. Jets retraction is well visible. By the end of the video it is possible to observe the beginning of RM734 crystallization.}
\movie{500$\mathrm{{\mu}m}$ RM734 droplet on a diffused doped LN substrate. Temperature ranges between 165 and 125°C with a cooling rate 0.2°C/s.}
\movie{250$\mathrm{{\mu}m}$ RM734 droplet on a diffused doped LN substrate. Temperature ranges between 125 to 65°C with a cooling rate of 0.07°C/s. Several instability events are observable after the first at T = 124°C. Interestingly, the droplet tends to regain a dome shape after each instability. By the end of the video it is possible to observe the beginning of crystallization.}
\movie{$1.1\times10^3\mathrm{{\mu}m}$ RM734 droplet on an undoped LN substrate. Temperature range: 133 – 123°C. Cooling rate 0.3°C/s. Several instabilities with fast jets ejection rapidly forming satellite small droplets are observable, the first one occurring at T = 131°C. The last shape instability is instead characterized by a violent ejection of fluid material, resulting in a kind of droplet destruction.}